\documentclass{article}[10pt]
\usepackage{setspace}
\usepackage{amsfonts}
\usepackage{amsmath}
\usepackage{graphicx}
\usepackage{amssymb}
\usepackage{amsmath}
\usepackage[pagewise]{lineno}
\usepackage{hyperref}

\input{tcilatex}
\begin{document}

\title{Turbulence phenomena for viscous fluids by a phase field model.
Vorticies and instability}
\author{Mauro Fabrizio$^{1}$ \and $^{1}$ Department of Mathematics,
University of Bologna, Italy}
\maketitle

\begin{abstract}
Through Ginzburg-Landau and Navier-Stokes equations, we study turbulence
phenomena for viscous incompresible and compressible fluids by a second
order phase transition. For this model, the velocity is defined by the sum of
classical and whirling components. Moreover, the laminar-turbulent
transition is controlled by rotational effects of the fluid. Hence, the
thermodynamic compatibility of the differential system is proved. 

This model can explain 
the turbulence by  instability effects motivated by a double well potential 
of the Ginzburg-Landau  equation.
The same model is used to understand the origins of tornadoes 
 and the birth of the vortices resulting from the fall of water in a
vertical tube. Finally, we demonstrate how the weak Coriolis force is able
to change the direction of rotation of the vortices by modifying the minima
of the phase field potential.
\end{abstract}


\section{Introduction}

Turbulence phenomena in a viscous fluid was for a long time a controversial
problem for the study and modeling of the transition from laminar flow to
turbulent behavior\cite{Ar-St},\cite{Ch},\cite{FABE},\cite{Ru}.\cite{St},%
\cite{Wi}. In many papers, it has been tried a modeling using only the
classical Navier Stokes equations. In other works, the turbulence is studied
by a stochastic cascade model. More recently, the turbulence is described by
a phase transition, but not always with a suitable connection with Navier
Stokes equations. In a previous work \cite{F2}, we have presented a
transition model by a system, where the Navier Stokes equation are
associated with the Ginzburg-Landau equation \cite{F1},\cite{landau},\cite%
{Lifs. Pit} and where the transition is described by a phase field $\varphi $%
\ and controlled not by the velocity $v$ of the fluid, but by its $%
\left\vert \nabla \times \mathbf{v}\right\vert $. This choice is motivated
by the results presented in many articles \cite{Go}, \cite{Hu},\cite{St}, 
\cite{Zu}, in which it is observed that the roughness of the walls or the
obstacles inside a channel can anticipate the transition to turbulence,
because they produce vortices. Indeed, it seems to us not convenient to
believe that a laminar flow in a pipe, consisting only of perfectly parallel
velocities, can be transformed into a turbulent flow, when the velocity
exceeds a given value predicted by the Reynolds number. On the other hand,
it seems reasonable to assume that an appropriate (even small) disorder flow
is needed for the transition.

In the paper we also want to emphasize, as the model proposed be convenient
to describe the instability evident in phenomena of turbulence transitions.
Unluckily, this phenomenon was not enough considered in literature. In this
paper, we recall the model of a viscous incompressible fluid studied in \cite%
{F2} with some corrections and improvements. In particular, we introduce a
small (but important) modification, because now the phase $\varphi \in
(-1,1) $, unlike of the previous work, where $\varphi \in \left[ 0,1\right) $%
. By this change, the instability effect is now described through the
bifurcation, which is manifest in the Landau potential \cite{landau}, when
the transition triggers. Because in such a case the potential goes from a
minimum to a double well potential.

Hence, in the paper we present an extension to compressible viscous fluids
and its thermodynamic compatibility. As in \cite{F2}, for these fluids we
suppose the threshold, that identifies the transition, given by a suitable
value of $\left\vert \nabla \times \mathbf{v}\right\vert $.

Moreover, there is a great similarity between turbulence phenomena for
viscous fluids and superfluids in Helium II (see \cite{Ar-St},\cite{F11},%
\cite{Kapit},\cite{landau},\cite{Lifs. Pit}\cite{Lond1},\cite{Mend1},\cite%
{Tisz}) and superconductivity \cite{G-L}.

Finally, as a consequence of this model, it follows that the differential
system can be well-posed problem only if we are in the laminar phase.
Otherwise, when we are in turbulence flow, the instability of the model
makes the system ill posed.

In the last part, we consider phenomena for which the transition does not
produce turbulent effects, because the vortices have ample dimensions as in
the tornadoes or in the fall of water in the hole of a sink. To describe
these new effects, we considered the same differential systems, but with a
different symmetry of Ginzburg-Landau potentials. \ So for these phenomena, the weak force of Coriolis plays an important role, naturally not generating the vortices, but indirectly influencing the direction of rotation, because it is able to modify the minima of the Ginzburg-Landau potential. 

\section{Incompressible fluids, turbulence and instable behavior}

In this paper, the mathematical model proposed for a turbulence flow in a
viscous fluid is confined in a smooth domain $\Omega \subset IR^{3}~$Hence,
we suppose the phenomenon consequence of a phase transition, such that the
transfer from laminar to turbulent flow is checked by a order parameter (or
phase field) $\varphi \in \left( -1,1\right) $.

Following $\left[ 1\right] $, the velocity\ $\mathbf{v}$ of fluid is
composed by the normal velocity $\mathbf{v}_{n}$ and the rotational
component $\mathbf{v}_{s}$. Moreover, for incompressible fluids $\mathbf{v}%
_{n}$ and $\mathbf{v}_{s}$ are related by the constraints ~%
\begin{equation}
\mathbf{v}_{s}\mathbf{=}\nu \mathbf{\nabla \times v}_{n}~,~~\nabla \cdot 
\mathbf{v}_{n}=0  \label{1}
\end{equation}%
with $\nu $ scalar and positive coefficient. Moreover, we assume the
velocity $\mathbf{v}$ given by the following relationship between $v_{n}$
and $v_{s}$ 
\begin{equation}
\mathbf{v}=\mathbf{v}_{n}+\varphi \mathbf{v}_{s}  \label{2}
\end{equation}%
moreover, the phase $\varphi \in \left( -1,1\right) ~$satisfies the
Ginzburg-Landau equation 
\begin{equation}
\rho _{0}\dot{\varphi}=\nabla \cdot L\nabla \varphi -NF^{\prime }(\varphi
)+\alpha G^{\prime }(\varphi )\mathbf{v}_{s}^{2}  \label{3}
\end{equation}%
where $\rho _{0}>0$ is the density of the incompressible fluid, $\alpha >0$
and $L(x)>0$ for all $x\in \Omega $. Besides, the two potentials $F$ and $G,$
are such that $F$ is given by a parabolic function with minimum for $\varphi
=0$ and $G$ by a function with two well potential and maximum in $\varphi
=0. $As \ an example we propose for $F$ and $G$ the functions 
\begin{equation}
F(\varphi )=\frac{\varphi ^{2}}{2},~~~~~G(\varphi )=\frac{\varphi ^{4}}{4}+b%
\frac{\varphi ^{3}}{3}-\frac{\varphi ^{2}}{2}-b\varphi ,~~~with~\varphi \in
\left( -1,1\right)  \label{4}
\end{equation}

Finally, the velocity $\mathbf{v}_{n}$ satisfies an extension of
Navier-Stokes equation%
\begin{equation}
\rho _{0}\mathbf{\dot{v}}_{n}=-\nabla p-\mu \nabla \times \nabla \times 
\mathbf{v}_{n}+\gamma \nabla \times (\dot{G}(\varphi )\mathbf{v}_{s})+\rho
_{0}\mathbf{b}  \label{5}
\end{equation}%
with the scalar $\gamma >0$ and the body force $b.$

Now, we introduce the boundary conditions on the smooth domain $\Omega \in
IR^{3}$ related with the system (\ref{3}-\ref{5}) on two domains $\partial
\Omega _{1}\neq 0$ and $\partial \Omega _{2}$

\begin{center}
$%
\begin{array}{c}
\left. \nabla \times \mathbf{v}_{n}\times \mathbf{n}\right\vert _{\partial
\Omega _{1}}=0 \\ 
\left. \mathbf{v}_{n}\right\vert _{\partial \Omega _{2}}=0%
\end{array}%
$
\end{center}

and 
\begin{equation}
\left. \nabla \varphi \cdot \mathbf{n}\right\vert _{\partial \Omega }=0
\label{7}
\end{equation}

In this section, we suppose in (\ref{4}) $b=0$, so we have that the
laminar-turbulent transition occurs when $(\frac{a}{N}v_{s}^{2}-1)$ changes
sign. Indeed, for $\frac{a}{N}v_{s}^{2}>1$, we are in turbulent phase.
Otherwise, if $\frac{\alpha }{N}v_{s}^{2}<1$ we are in laminar flows.

\begin{center}
\includegraphics[scale=0.4]{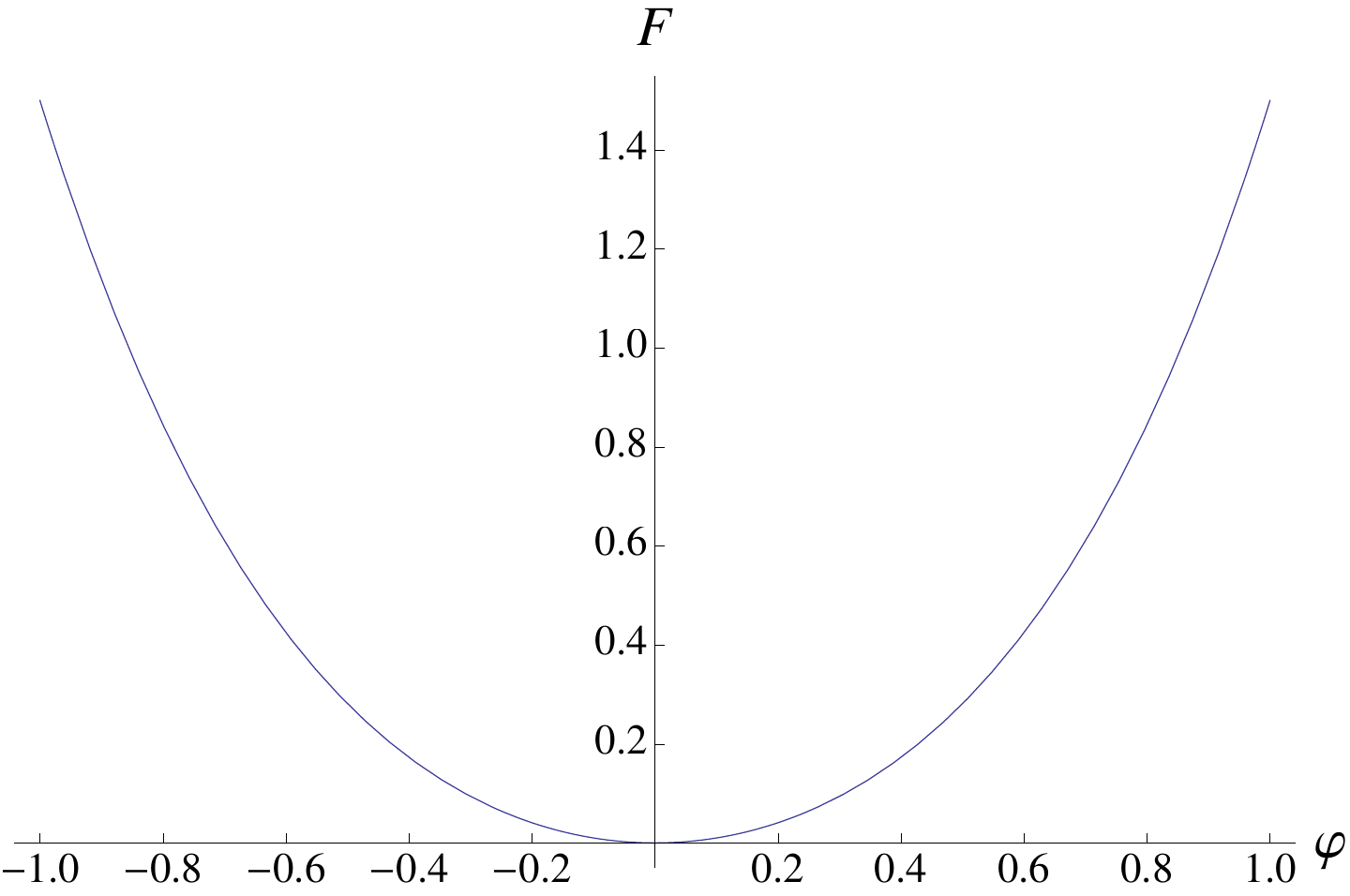} %
\includegraphics[scale=0.4]{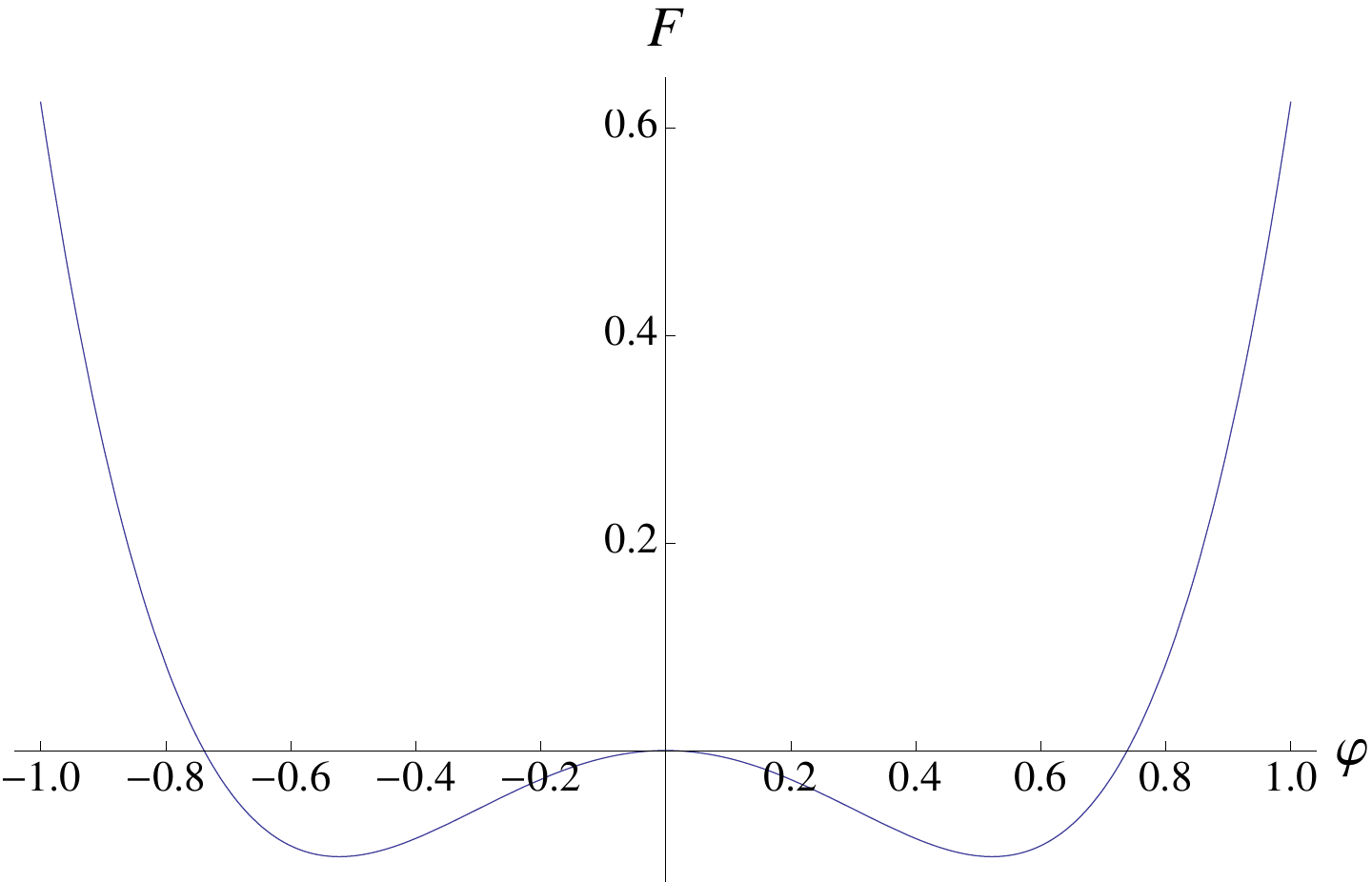}

Fig.1 - The two pictures describe the potential $F(\varphi )$ of (\ref{4})
the first and the potential $G(\varphi )$ with $b=0~\ $the second.
\end{center}

The turbulence behavior is related with the double well function of$%
~G(\varphi )$ $~$of (\ref{4}), because the two minima are to the same
height. So that, in such a case the system is unable to select between the
two minima. So in any point, we can have a different choose of the minimum.
So this instability produces a turbulence behavior of the fluid.

Even if our system can be instable, anyway we \ are always able to obtain
the compatibility of the system with Thermodynamics, which for isothermal
processes assumes the form of Dissipation Principle \cite{F1},\cite{frem},%
\cite{Fr-Gur}:

There exists a state function $\psi (\varphi ,\nabla \varphi ),$~called
internal energy, such that:%
\begin{equation}
\rho _{0}\dot{\psi}(\varphi ,\nabla \varphi )\leq \mu (\nabla \times \mathbf{%
v)}^{2}+\rho _{0}\dot{\varphi}^{2}+\frac{1}{2}\frac{d}{dt}(L(\nabla \varphi
)^{2}+2N\ F(\varphi ))  \label{8}
\end{equation}

From the system (\ref{3}-\ref{5}), with the restriction (\ref{1}-\ref{2})
and the boundary conditions (\ref{6}-\ref{7}), we obtain the internal
mechanical and structural power respectively%
\begin{equation*}
P_{m}^{i}=\mu (\nabla \times \mathbf{v}_{n})^{2}+\alpha \dot{G}(\varphi )%
\mathbf{v}_{s}^{2}
\end{equation*}%
\begin{equation*}
P_{s}^{i}=\rho _{0}\dot{\varphi}^{2}+(\frac{L}{2}(\nabla \varphi
)^{2})^{\cdot }+N~\dot{F}(\varphi )-\alpha \ \dot{G}(\varphi )\mathbf{v}%
_{s}^{2}
\end{equation*}

Finally, by Dissipation Principle%
\begin{equation*}
\rho _{0}\dot{\psi}(\varphi ,\nabla \varphi )\leq P_{m}^{i}+P_{s}^{i}
\end{equation*}%
we obtain the inequality (\ref{8}) with free energy 
\begin{equation*}
\psi (\varphi ,\nabla \varphi )=\frac{1}{2\rho _{0}}(L(\nabla \varphi
)^{2}+2N\ F(\varphi )).
\end{equation*}

\section{A generalized Navier-Stokes and Ginzburg-Landau equations for
turbulence compressible fluids}

As in the previous section, we suppose the velocity of a compressible fluid
composed by a normal velocity $v_{n}$ and the rotational component $v_{s}$
by the same equation (\ref{2}), while the equation (\ref{1}) is replaced by
the following 
\begin{equation}
\rho (x,t)\mathbf{v}_{s}(x,t)=\tilde{\lambda}\nabla \times \rho (x,t)\mathbf{%
v}_{n}(x,t)  \label{11}
\end{equation}%
where $\rho $ denotes the density of the fluid and $\tilde{\lambda}$ a
positive coefficient. So, from (\ref{11}) we \ obtain%
\begin{equation*}
\nabla \cdot (\rho \mathbf{v}_{s})=0
\end{equation*}%
so from continuity equation and by the equation (\ref{2}) 
\begin{equation}
\frac{\partial \rho }{\partial t}=-\nabla \cdot (\rho \mathbf{v)=-}\nabla
\cdot (\rho \mathbf{v}_{n}\mathbf{)-}\rho \mathbf{v}_{s}\cdot \nabla \varphi
\label{12}
\end{equation}%
hence, we have the motion equation 
\begin{equation}
\rho \mathbf{\dot{v}}_{n}=-\nabla (p(\rho )+(2\mu +\lambda )\nabla \cdot
\nabla \mathbf{v}_{n}-\mu \nabla \times \nabla \times \mathbf{v}_{n}+\gamma
\nabla \times (\rho \dot{G}(\varphi )\mathbf{v}_{s})+\rho \mathbf{b}
\label{13}
\end{equation}%
where $\mu $~and $\gamma $ are positive constants, such that $\lambda \gamma
=\alpha $ and $\mathbf{b}$ the body force.

Finally, as in the previous section, we consider the Ginzburg-Landau equation%
\begin{equation}
\rho \dot{\varphi}=\nabla \cdot \rho L\nabla \varphi -\rho NF^{\prime
}(\varphi )+\alpha \rho G^{\prime }(\varphi )\mathbf{v}_{s}^{2}  \label{14}
\end{equation}%
where the coefficients $N,L$and the potentials $F(\varphi )$ and $G(\varphi
) $ are the same of the previous section, defined by (\ref{4}).

So, the internal power $P_{m}^{i},P_{s}^{i}~$are given by%
\begin{equation*}
P_{m}^{i}=\mu (\nabla \times \mathbf{v}_{n})^{2}+(2\mu +\lambda )(\nabla ^{2}%
\mathbf{v}_{n})^{2}+\alpha \rho \dot{G}(\varphi )\mathbf{v}_{s}^{2}
\end{equation*}

\QTP{Body Math}
\begin{equation*}
P_{s}^{i}=\rho \dot{\varphi}^{2}+(\rho \frac{L}{2}(\nabla \varphi
)^{2})^{\cdot }+\rho N~\dot{F}(\varphi )-\alpha \ \rho \dot{G}(\varphi )%
\mathbf{v}_{s}^{2}
\end{equation*}

Hence, the internal energy $\psi (\varphi ,\nabla \varphi )$~satifies the
inequality%
\begin{equation}
\rho \dot{\psi}(\varphi ,\nabla \varphi )\leq P_{m}^{i}+P_{s}^{i}=
\label{15}
\end{equation}%
\begin{equation*}
=\mu (\nabla \times \mathbf{v}_{n})^{2}+(2\mu +\lambda )(\nabla ^{2}\mathbf{v%
}_{n})^{2}+\rho \dot{\varphi}^{2}+\rho \frac{d}{dt}(\frac{L}{2}\nabla
\varphi )^{2}+N~F(\varphi ))
\end{equation*}

\section{ Vorticies and tornadoes}

In this last section, we observe how the same model considered in the sect.
2, can study new phenomena like tornadoes or water vorticity, when it falls
into a vertical deep hole. For these models, we consider the equations (\ref%
{1}-\ref{3}) and (\ref{4}) with $b\mathbf{\neq }0.$

\paragraph{a .Water vorticity\protect\bigskip}

It is known that usually the water that descends into a sink or more
generally into a eddy, rotates in one direction in the northern hemisphere
and in the other direction on the south hemisphere. In some papers, this
different behavior (as well as for tornatoes) is explained by the Coriolis
force. However, it is often observed that this opposite trend can not be
directly attributed to the Coriolis force, since the rotation speed of the
earth (hence the Coriolis force) is very weak. Otherwise in our model, this
force can work indirectly by the coefficient $b$ of $G(\varphi )$ in (\ref{4}%
). So that, a suitable definition of $b$ could be

\begin{equation}
b=\tau \mathbf{\omega \times v}_{r}\cdot \mathbf{t}  \label{17}
\end{equation}%
where $\mathbf{t}$ is the tangent to the parallel of the earth and $\mathbf{v%
}_{r}$ is the velocity relative to the terrestrial surface, while $\mathbf{%
\omega }$ is the angular velocity of the earth and $\tau $ an appropriate
constant coefficient.

In fact, to explain the rotation of the fluid that is generated during the
fall in a vertical channel, we can use the equation (\ref{2}) associated
with (\ref{1}) and (\ref{3}) with the potential $F(\varphi )$ and $G(\varphi
)$ of (\ref{4}), where $b\neq 0$ can be defined by (\ref{17}).

Now we consider the trajectory of a water particle when it enters the sink
tube. Then, if the speed is small enough the motion is laminar. So the water
falls inside the tube without any rotational component. If instead the speed
is sufficiently high, then we observe, as as a result of equation (\ref{2}),
that the motion does not remain on the same vertical plane, but a rotational
component is generated around the tube axis. In other words, this effect is
due to variation in speed when water enters the pipe. In fact, because of
the trajectory we have a rotary motion with axis tangent to the edge of the
tube, which on the basis of equation (\ref{2}) generates a new rotational
component $\varphi \mathbf{v}_{s}$ around the tube axis, whose rotation
sense is consequent to the sign of $\varphi $.Therefore, it is not the
direct action of the Coriolis force that causes the variation of the water
rotation in the two hemispheres, but the structural change caused by the
transition described by eq. (\ref{3}), which modifies the sign of $b$ of the
eq. (\ref{4}) and therefore the absolute minimum $\varphi _{m}$ of the
potential (\ref{4})$_{2}$  changes with the sign of $b$, which is
related with the angular velocity $\mathbf{\omega }$, which conditions the
rotation of water according to the hemisphere in which the phenomenon occurs.

Therefore, according to the sign of $b$, we obtain the following two graphs.

\begin{center}
\includegraphics[scale=0.4]{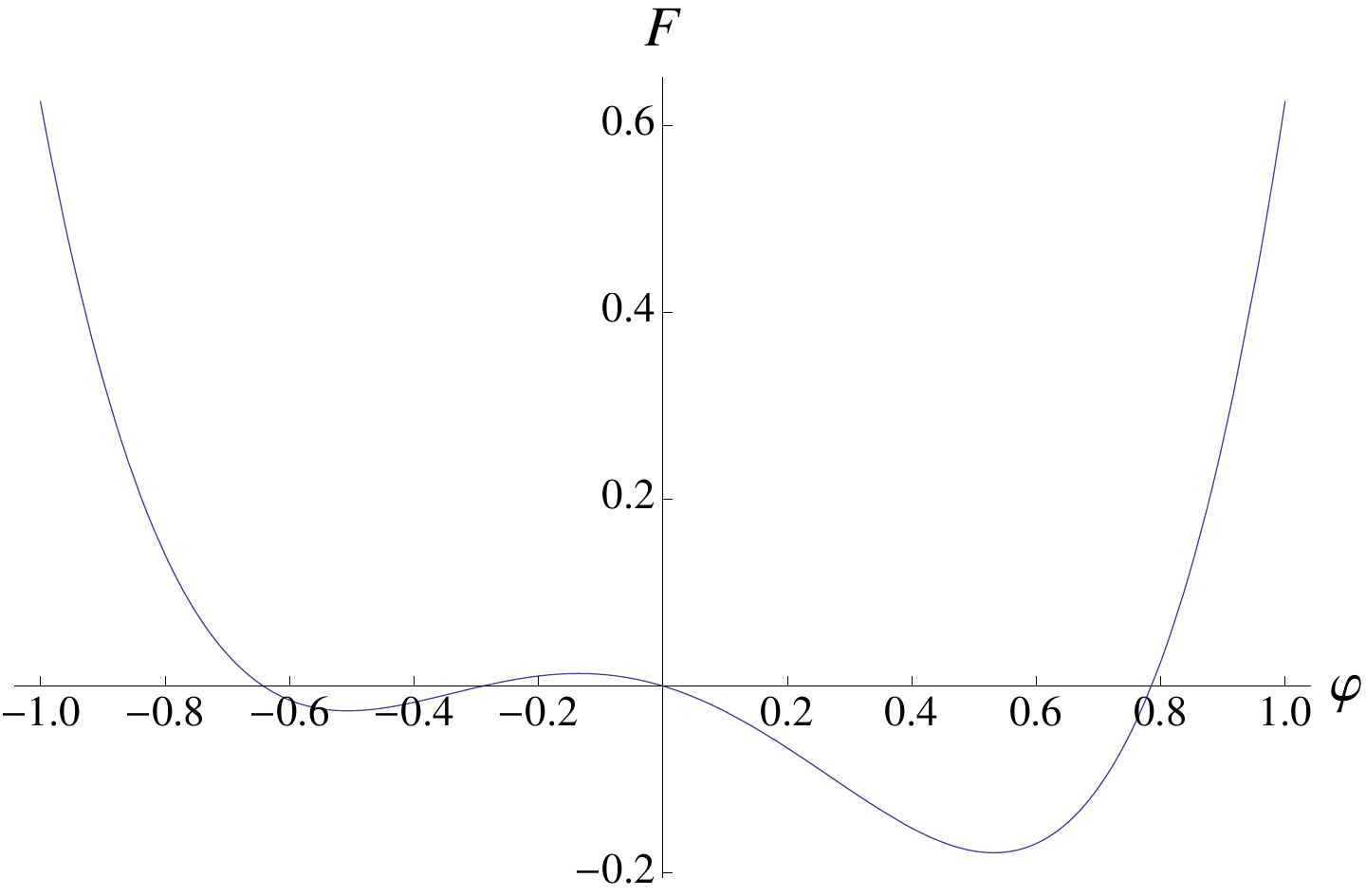} %
\includegraphics[scale=0.4]{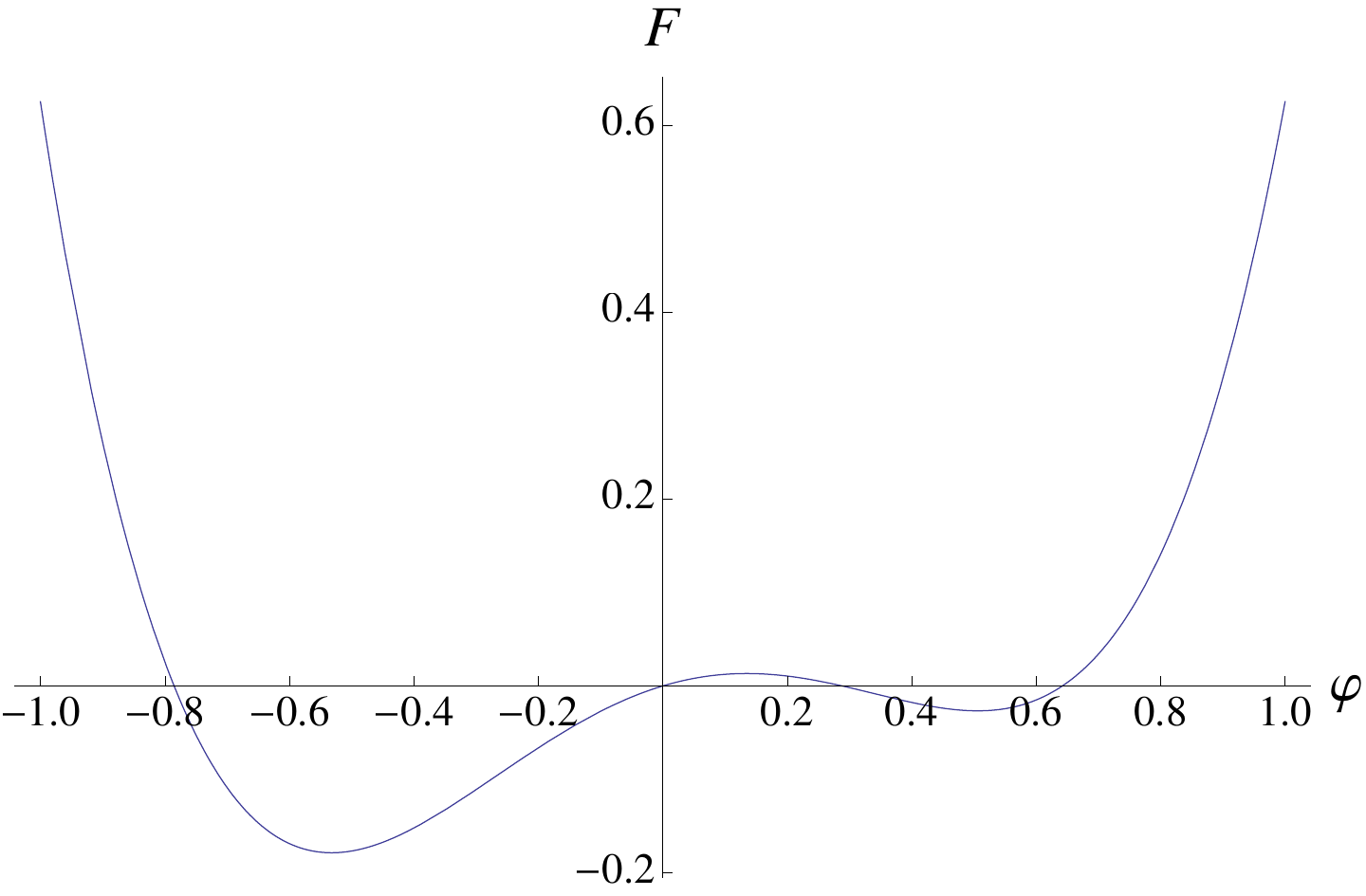}

Fig.2 - The two pictures are obtained with two different value of the
coefficient $b.~$In one we have $b>0$ and in the second $b<0.$
\end{center}

\paragraph{b -Tornadoes}

Then as a consequence of the wind action, we suppose to have in the
atmosphere clouds vortices. So, when the vorticity $\mathbf{v}_{s}=\nu
\nabla \times \mathbf{v}_{n}$~is quite high and its modulus exceeds the
critical value, as a coseguence of equation (\ref{2}) we observe that the
component $\varphi \mathbf{v}_{s}$ of the velocity is normal to vortices.
Then, it begin the tornado effect towards the earth.

It should be noted that this model is able to explain the rotation of the
tornado vortexes in the two hemispheres of the earth. In fact, to explain
this different behavior we have to use again the coefficient $b$ defined in (%
\ref{17}). In fact, when $b$ is positive, the graph of the function $%
G(\varphi )$ in (\ref{4})$_{2}$ has an absolute minimum for $\varphi >0$ and
therefore a direction at the term $\varphi \mathbf{v}_{s}$, if instead $b<0$
we have the absolute minimum for $\varphi <0$ and therefore an opposite
direction to the previous one.

Finally, in agreement with the sign of $b$, we have two similar graphs to
the Fig. 2.

\end{document}